\documentclass[11pt,preprint,graphicx]{aastex}

\usepackage[fleqn]{amsmath} 
\usepackage{amsfonts} 
\usepackage{amstext}
\usepackage{amssymb} 
\usepackage[colorlinks=true, citecolor=blue, linkcolor=blue, breaklinks=true]{hyperref}
\usepackage{natbib}
\usepackage{graphicx}

\def\etal{\it et al. \rm }

\begin{document}

\title{Anomalous Stellar Populations in LSB Galaxies}

\author{James Schombert$^{1}$ and Stacy McGaugh$^{2}$}
\affil{$^{1}$Institute for Fundamental Science, University of Oregon, Eugene, OR 97403}
\affil{$^{2}$Department of Astronomy, Case Western Reserve University, Cleveland, OH 44106}

\begin{abstract}

\noindent We present new HST WFC3 near-IR observations of the CMD's in two LSB
galaxies, F575-3 and F615-1, notable for having no current star formation based on a
lack of H$\alpha$ emission.  Key features of the near-IR CMD's are resolved, such as
the red giant branch (RGB), the asymptotic giant branch (AGB) region and the top of
the blue main sequence (bMS).  F575-3 has the bluest RGB of any CMD in the
literature, indicating an extremely low mean metallicity.  F615-1 has unusually wide
RGB and AGB sequences suggesting multiple episodes of star formation from metal-poor
gas, possibly infalling material.  Both galaxies have an unusual population of stars
to the red of the RGB and lower in luminosity than typical AGB stars.  These stars
have normal optical colors but abnormal near-IR colors.  We suggest that this
population of stars might be analogous to local peculiar stars like Be stars with
strong near-IR excesses owing to a surrounding disk of hot gas.

\end{abstract}

\section{Introduction}

The advent of space imaging allows the direct inspection of the stellar populations
in nearby galaxies.  Depending on their distance, the resolved stellar populations
can reach below the red clump and into the main sequence (Weisz \etal 2019), regions
of fundamental interest in untangling the star formation history of galaxies.  
More distant galaxies, outside the Local Group, we must be satisfied with information
about the high luminosity portion of a galaxy's color-magnitude diagram (CMD), the
RGB, the region of AGB stars and the top of the main sequence.  Many of these stars
either have short lifetimes (i.e., the top of the main sequence) or have recently
left the main sequence and RGB (i.e., AGB stars).  Both types are of high
interest in galaxies with some evidence of recent star formation (blue colors or
H$\alpha$ emission).

Optical CMD's provide a sharp view of the early phases of star formation (SF): the
top of the main sequence and the blue helium burning branch (the so-called blue
plume).  A rule of thumb is that resolution of approximately two magnitudes below the
tip-of-the-RGB (TRGB) allows determination of the last Gyr of star formation using
CMD-fitting algorithms such as MATCH (Dolphin 2002, see Weisz \etal 2015).  Near-IR
CMD's provide a longer timescale view of the star formation history of a galaxy with
more information on the AGB population (with lifetimes in the 3 to 5 Gyrs range).  In
addition, the position of the RGB and AGB stars is a strong function of metallicity,
allowing for some information on the chemical enrichment history of a galaxy to be
deduced.

This study is a continuation of HST imaging of a subset of the low surface brightness
(LSB) catalog of Schombert \etal (1997).  Late-type LSB galaxies are very blue, an
indicator of recent star formation and/or very low metallicities.  Current star
formation is not uncommon in LSB galaxies (Schombert, McGaugh \& Maciel 2013)
although typically at very low levels to match their low stellar densities.  In
parallel, low metallicity is presumed as their low stellar densities plus low current
SFR implies a suppressed history of star formation and, therefore, slow chemical
evolution.  However, LSB galaxies form a surprisingly tight correlation between total
stellar mass and their current SFR (the so-called main sequence for star-forming
galaxies, Noeske \etal 2007, Speagle \etal 2014, McGaugh, Schombert \& Lelli 2017)
implying a steady rate of star formation averaged over a Hubble time.

As can be seen in Figure 1 of Schombert, McGaugh \& Lelli (2019), the collection of
low-mass HSB and LSB galaxies form a well-defined sequence just slightly to the right
of the line of constant star formation.  Thus, the current SFR in low-mass galaxies
is fairly similar to the average SFR over a Hubble time and excludes a scenario where
the past SFR is much higher (or lower) as this would dramatically over or under
produce the final stellar mass of the galaxy (see also Kroupa \etal 2020).  One could
imagine a much higher past SFR with a later epoch of initial star formation, but this
goes against the evidence of old ($\tau$ $>$ 10 Gyrs) stars in observed CMD's
(Schombert \& McGaugh 2014).  The simplest monotonic function that reproduces the
main sequence is a weakly declining exponential SFH with very little room for large
deviations from this history, such as strong bursts in recent epochs, without wildly
different observed optical and near-IR colors (see Figure 3, Schombert, McGaugh \&
Lelli 2019).

This proposed smooth SFH is in strong contrast to the observed SFH in many nearby
dwarfs as deduced from their resolved CMD's (Weisz \etal 2011).  Global colors, color
gradients and SED modeling can only constrain a galaxy's SFH to a certain degree.  A
much sharper view is obtained by resolving the underlying stellar populations, even
just the upper portion of the CMD.  Using optical CMD's, Weisz \etal (2014) decoded
the SFH of 40 nearby Local Group dwarfs and concluded that an exponentially declining
SFR is a good match to most dwarfs.  However, the fraction of stellar mass produced
as a function of time varied significantly from low-mass to high-mass dwarfs.
Several dwarfs also display long quiescent epochs between stronger than average
bursts, i.e. not the smooth, monotonic exponential one presumes.

Imaging in the near-IR also provide a view into the chemical history of a galaxy.
Metallicity serves as a secondary clock of star formation as enrichment is a
relatively rapid process even for galaxies with lower SFRs.  Decreasing the age of a
stellar population to increase its mean SFR would drive the chemical enrichment to
higher final values.  The mean metallicity of LSB galaxies either through O/H
determination (McGaugh 1994) or direct measures of their rHeB branches (Schombert \&
McGaugh 2015) are extremely low ([Fe/H] $<$ $-$1.5) and argue against periods of
strongly elevated star formation in the past.  Their high gas fractions indicates
that metal-poor gas is still within the galaxy boundaries and unprocessed into stars.
Recent burst epochs can be reconciled with low current [Fe/H] values if these bursts use
part of this unprocessed gas supply.

For this study, we have selected two seemingly opposite LSB galaxies in their optical
and near-IR colors, yet with similar stellar masses and gas fractions.  In addition,
both have no detectable H$\alpha$ emission signaling current SF (i.e., they are
currently in a quiescent phase).  By selecting two quiescent galaxies with very
different colors, we hope to untangle their SFH and apply this knowledge to a wider
spectrum of LSB galaxy colors.  We have selected near-IR imaging as the addition
bonus of investigating the proportion of asymptotic giant branch (AGB) stars in LSB
galaxies, a key component to modeling their $M/L$ values for studies of the baryonic
mass in galaxies.

\section{Data}

The two galaxies chosen for near-IR HST imaging are F575-3 (also known as KDG215 and
D575-3) and F615-1 (also known as KDG023).  Both galaxies were cataloged (Schombert
\etal 1992) during a visual search of PSS-II IIIaJ plates (Reid \etal 1991) on the
Palomar 48-inch Schmidt telescope between 1987 and 1989.  The goal of that catalog
was to increase the number of LSB galaxies under the UGC criterion of a one arcmin
diameter given the increased surface brightness sensitivity of the latest Kodak blue
emulsions.  Discovery was by a simple eyepiece inspection with cross reference to the
transparent overlays produced by the Smithsonian Astrophysical Observatory that  
labeled all NGC, IC and UGC objects on each Schmidt plate (see Schombert \etal 1997).
A previous search of Sculptor region included a handful of the Palomar LSB dwarfs
(Karachentseva 1968) which were re-discovered (not an uncommon event for LSB
galaxies).  As we have published several follow-up papers on these objects without
knowing their KDG designations, we will continue to use their Palomar designations.

The visual appearance of both galaxies on the IIIaJ plates (see Figures 1 and 2)
satisfied a `dI' classification.  This was owing to a smooth, exponential shape to the
objects, no indication of HSB knots (typical of dIrr galaxies) or disk-like
morphology (dS).  The dI's are distinguished from dE by being less elliptical in
their isophotes and not as centrally concentrated, for example a core that is offset
from their outer isophotes.  Ultimately, the detection of HI at Arecibo was the final
criteria for the dI classification.

Past observations for F575-3 and F615-1 are summarized in Table 1.  HI detection was
obtained in Eder \& Schombert (2000), updated ALFALFA observations gives HI fluxes of
5.51$\pm$0.06 and 2.46$\pm$0.05 Jy respectfully (Haynes \etal 2018).  Optical and
H$\alpha$ imaging was obtained on the KPNO 2.1m (Schombert, McGaugh \& Maciel 2013).
$Spitzer$ 3.6$\mu$m imaging was acquired for both galaxies during cycle 25 (Schombert
\& McGaugh 2014) with total 3.6 magnitudes of 14.40$\pm$0.08 and 13.48$\pm$0.05.
Using the distances determined by the TRGB method (see \S4), these luminosities map
into gas and stellar masses as given by the prescriptions outlined in Schombert,
McGaugh \& Lelli (2020) and listed in Table 1.

\begin{figure}
\centering
\includegraphics[scale=0.35,angle=0]{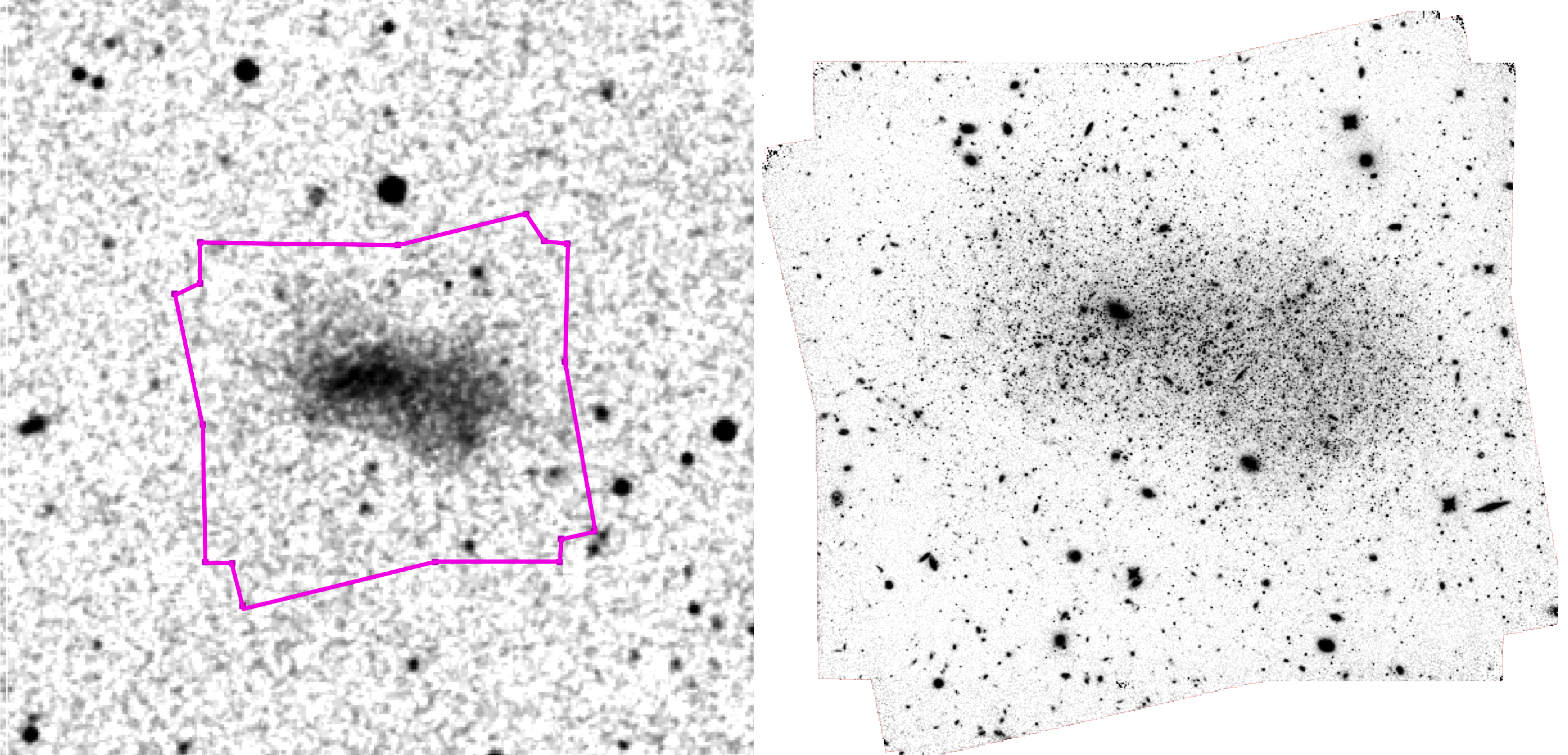}
\caption{\small The discovery image for F575-3 from the Second Palomar Sky Survey
IIIaJ plates is shown on the left.  The PSS-II image is 4.5 arcmins across with a
magenta outline that displays the HST WFC3 image shown to the right.  The WFC3 images
have an effective exposure time of 5,224 secs through the F110W filter with a plate
scale of 0.13 arcsecs per pixel.  Many of the knots near the galaxy core are, in
fact, objects in a background group of galaxies and not associated with the target
galaxy.  As none of the gas-rich objects in the background group were detected in our
Arecibo observations out to 25,000 km/sec, this would place the group beyond 300 Mpc.  
All the brightest point sources were confirmed visually to be stellar.
}
\label{f575-3}
\end{figure}

\begin{figure}
\centering
\includegraphics[scale=0.35,angle=0]{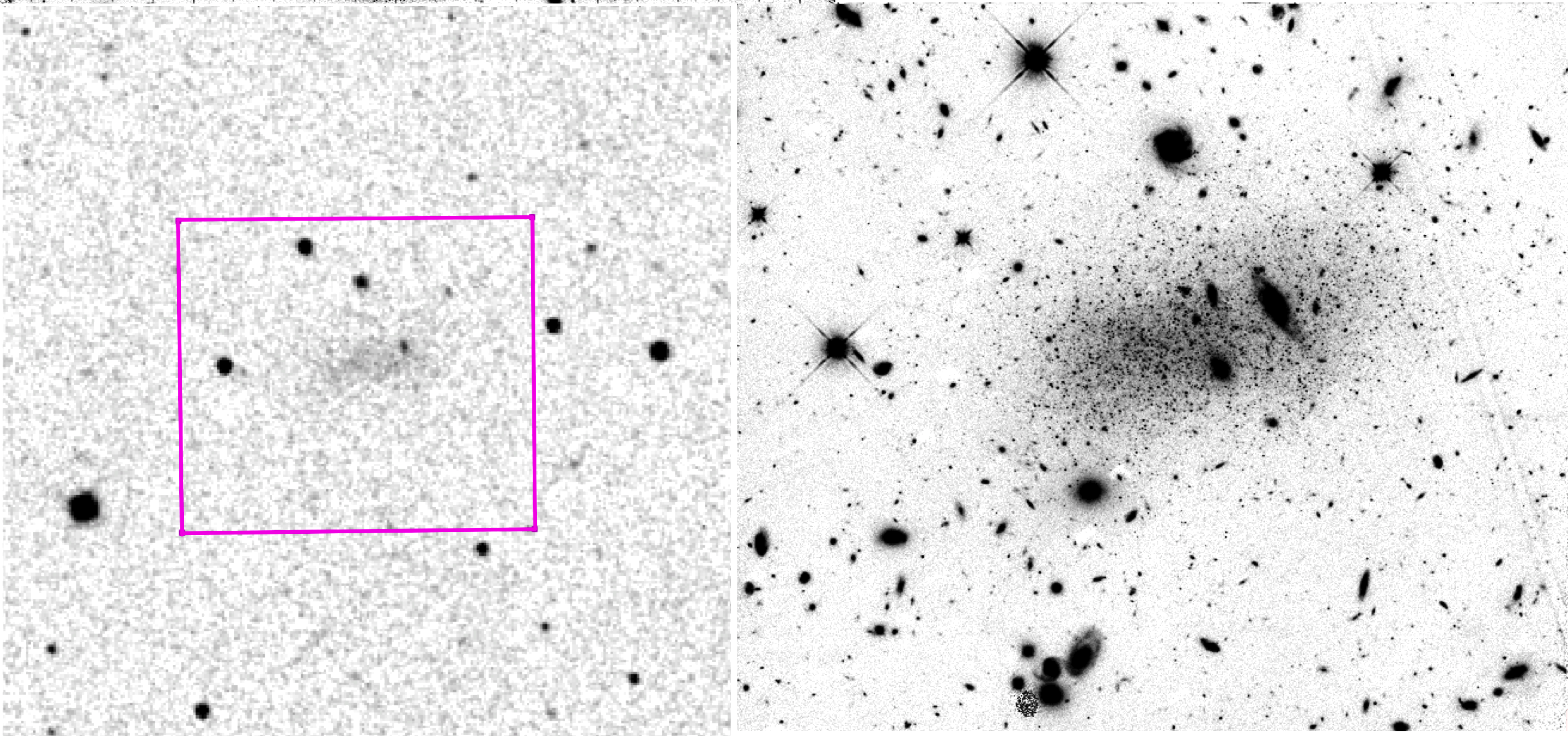}
\caption{\small Same as Figure 1 for F615-1.  The discovery PSS-II image on left
with a width of 4.5 arcmins, WFC3 image on the right.  WFC3 exposure times and plate scale are identical to
F575-3.  Unfortunately, as with F575-3, the knots inside F615-1 are primarily owing to
a small group of galaxies behind the target and were masked from the photometry.
}
\label{f615-1}
\end{figure}

For the distances listed in Table 1, the H$\alpha$ imaging from Schombert, McGaugh \&
Maciel (2013) would have detected even a single OB HII region.  There are several LSB
galaxies in that sample with H$\alpha$ fluxes corresponding to SFR at, or below, the
$10^{-4}$ $M_{\odot}$ yr$^{-1}$ level, thus, we believe any current SF would have
been detected.  From their stellar masses, the main sequence for LSB galaxies
predicts between $10^{-3}$ and $10^{-4}$ $M_{\odot}$ yr$^{-1}$ of star formation
(i.e., between one OB association and an Orion complex level event).  In addition to
the lack of H$\alpha$, no star clusters were detected in the WFC3 images and any
optical knots from ground-imaged turned out to be background galaxies in the HST
imaging.  Our conclusion is that there is no recent SF ($\tau <$ 50 Myrs) in these
two dwarfs, despite their extremely high gas fractions.

F575-3 and F615-1 are similar in morphology and gas fraction (plus lack of H$\alpha$
emission), but differ in their optical to near-IR colors. Whereas, F615-1 has the
colors typical to an early-type spiral (see Table 1 and Figure 2 from Schombert,
McGaugh \& Lelli 2019), F575-3 is the bluest galaxy in the LSB catalog with a $B-V$
color of 0.26, similar to the blue colors of a starburst irregular.  However, despite
its red color, F615-1 has a high gas fraction indicating it is not a quenched dI. F575-3
has an even higher gas fraction making its lack of current SF truly anomalous among
late-type LSB dwarfs.


\begin{deluxetable}{lcc}
\tablecolumns{3}
\small
\tablewidth{0pt}
\tablecaption{Galaxy Characteristics}
\tablehead{
\colhead{} & \colhead{F575-3} & \colhead{F615-1}
}
\startdata
RA (J2000) Dec & 125541.0$+$191233 & 024325.5$+$164400 \\
$M_{3.6}$ & $-$14.1 & $-$16.6 \\
$M_*$ ($M_{\odot}$)$^a$ & 4.5$\times$$10^{6}$ & 4.5$\times$$10^7$ \\
$M_{gas}$ ($M_{\odot}$)$^b$ & 4.8$\times$$10^7$ & 9.1$\times$$10^7$ \\
$f_g$ & 0.91 & 0.67 \\
$R_{24}^c$ & 38 arcsecs (0.9 kpc) & 81 arcsecs (4.2 kpc) \\
B-V & 0.26 & 0.78 \\
V-3.6 & 1.25 & 3.51 \\
D (Mpc) & 5.1$\pm$0.2 & 10.6$\pm$0.5 \\
\enddata
\tablenotetext{a}{\ 3.6 luminosities and sizes from Schombert \& McGaugh (2014)}
\tablenotetext{b}{\ HI fluxes from Arecibo observations, Eder \& Schombert (2000)}
\tablenotetext{c}{\ semi-major axis at 24 $K$ mag arcsecs$^{-2}$}
\end{deluxetable}

\section{F110W/F160W Imaging}

Images for this project were obtained with HST using the Wide Field Camera 3 (WFC3).
The WFC3 is a fourth-generation imaging instrument with a near-IR channel that uses a
1k$\times$1k HgCdTe array with a pixel scale 0.135$\times$0.121 arcsec/pix and a
field of view 136$\times$123 arcsec.  Our exposures used the F110W and F160W filters
located inside the cold shield, which are the nearest equivalents to the traditional
Johnson $J$ and $H$.

Eight orbits were used in early 2018 for program HST-GO-15427 (PI: Schombert), four
orbits for each galaxies, two orbits per filter.  A four point dither was used to
improve resolution and eliminate cosmic rays.  Due to the low stellar density nature
of the targets, point source crowding was not a problem.  The resulting raw FLC
frames were combined using the AstroDrizzle package (Gonzaga 2012), which aligns the
images, identifies any additional cosmic rays, removes distortion, and then combines
the images after subtracting the identified cosmic rays.  The output DRC FITS images
are used for position reference, but photometry is performed on the raw FLC images.

Photometry of the drizzled frames was performed using the DOLPHOT package (Dolphin
2000) and the WFC3 modules.  DOLPHOT performs point-spread function (PSF) fitting on
all of the flat-fielded and CTE corrected images per field simultaneously. A refinement
of the shifts between the WCS of the observations, scale, and rotation adjustments is
done by DOLPHOT after a first estimate of these tasks is done by AstroDrizzle.  Some
manual masking to background was applied, but otherwise the standard techniques and
parameters were used including distortion and encircled energy corrections plus the
new WFC3 zeropoints.  Luminosities are expressed in the HST VEGAMAG system, and
comparison to models or simulations was made converting models into the HST filter
system.

The detection threshold was set for 3$\sigma$ above local sky.  Sources within 50
pixels of the edges were rejected, as were sources that overlapped within three
pixels.  Given these conditions, there were 5,350 point sources for F575-3 and 2,462
for F615-1, which maps well into the expected surface brightness differences from
their $Spitzer$ images.  Errors in $m_{F110W}$ varied from 0.01 at 24 mags to 0.25 at 27
mags.  Equivalent errors are found one magnitude brighter in F160W.  Incompleteness
was primarily due to background galaxies that account for 5 to 8\% of the field of
view.  Completeness was 100\% down to 25.5 F160W mags, 70\% by 26.0.

\begin{figure}[!t]
\centering
\includegraphics[scale=0.80,angle=0]{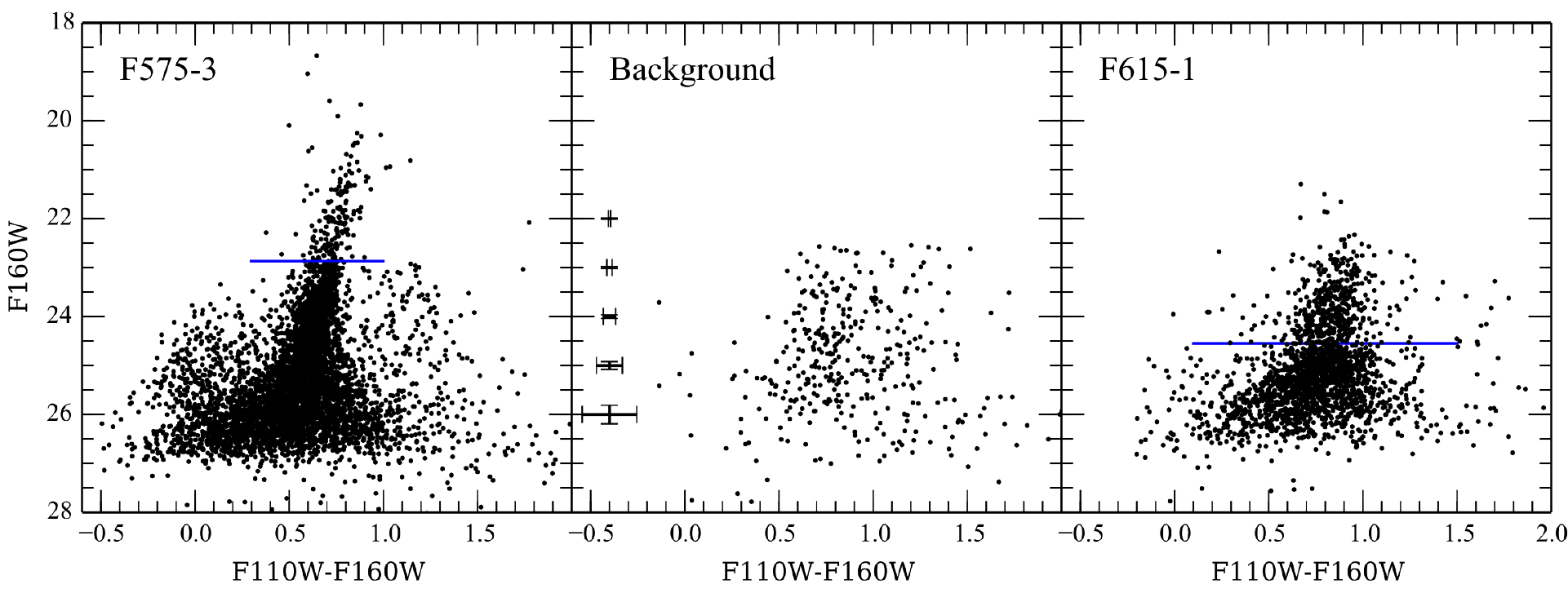}
\caption{\small The F110W-F160W color-magnitude diagrams for LSB dwarfs F575-3 and
F615-1.  The TRGB, estimated by an edge detection filter, is marked by the blue line.
F575-3 displays the features of a young dwarf population with a narrow RGB, strong
AGB branch, weak rHeB feature and numerous blue main sequence (bMS) stars.  F615-1
displays the features of an old dwarf with an unusually wide RGB and corresponding
wide AGB branch.  No obvious rHeB feature and few bMS stars.  The middle panel
displays the background/foreground population estimated from the outer edges of the
F615-1 frames.  Error bars in F160W and
F110W-F160W color are shown in middle panel.  }
\label{cmd}
\end{figure}

The resulting CMD's, corrected for Galactic extinction, are shown in Figure
\ref{cmd}.  Given the similarity in global characteristics, both galaxies have high
gas fractions, irregular morphology and no detectable current SF, the dichotomy in
their CMD's is striking (although less surprising when their differences in optical
colors is considered).  F575-3 displays all of the features of a star-forming dwarf
galaxy with a prominent blue and red helium branches (bHeB and rHeB), strong M-type
AGB colony and a dominant blue main sequence (bMS).  F615-1, in contrast, is
deficient in bHeB stars with a weak bMS population and few young AGB's.  F615-1 also
has unusually wide RGB and AGB branches.

Correction for Milky Way foreground stars and background distant galaxies is
problematic as both galaxies fill the WFC3 frame (F615-1 less so than F575-3 due to
its higher distance).  F575-3 has a higher central surface brightness and, although
smaller in isophotal radius than F615-1, has a significant population to the edge of
the frame.  Examining just the 20\% of the frame around the edges produces a CMD with
all same features of the core region, a distinct RGB and clear TRGB gap.  F615-1 is
lower in mean surface brightness, but more extended in size.  Dividing the F615-1
frame into elliptical area that contains the 23 $K$ mag arcsec$^{-2}$ isophote
results in a nearly 50/50 division of pixels.  We display the outer region as the
background CMD in middle panel of Figure \ref{cmd}.  As both F575-3 and F615-1 are of
high galactic latitude, foreground contamination should be similar.  

While there is a suggestion of a RGB in the foreground/background CMD, the
distribution is more in alignment with a Milky Way M dwarf population that range in
$J-H$ colors from 0.5 to 1.0.  Redder objects are probably high-redshift background
galaxies that have $J-H$ colors ranging from 0.8 to 1.5 (see Cowie \etal 1994).
While a majority of distant galaxies at these apparent magnitudes are resolved, and
thus eliminated by the photometry algorithms, a few compact galaxies would enter into
the sample.  The number of foreground stars with $J-H$ colors less than 1.0 in the
background CMD is slightly higher than expected from comparison star counts as a
function of apparent magnitude (Robin \etal 2003) which would seem to indicate that
even at large radii there still are detectable RGB stars in the frame.  The redder
than RGB color objects will be discussed in \S9.

The foreground corrected CMD's are very much in alignment with their global optical
colors, with F575-3 being extremely blue in both $B-V$ and $V-3.6$, while F615-1 has
colors typical of an early-type spiral.  The integrated $J-H$ colors deduced from the
CMD's is also in agreement with their measured $V-3.6$ colors.  The estimated TRGB
magnitude is shown as a blue line in Figure \ref{cmd} and discussed in the next
section.

\section{Distance}

F575-3 and F615-1 were selected for HST imaging with the expectation that both were
near the edge of useful CMD analysis.  Both were believed to have distances of 9 to
10 Mpc based on HI determined velocities compared to a simple Virgo infall model.
F575-3 has a CMB3K velocity of 716 km sec$^{-1}$ (which resolves into a distance of
9.9 Mpc for $H_o=75$); however, CosmicFlows-3 (Tully \etal 2019) gives a distance of
4.1 Mpc.  F615-1 has a CMB3K velocity of 595 km sec$^{-1}$ (which resolves into a
distance of 7.9 Mpc); however, CosmicFlows-3 gives it a distance of 12.1 Mpc.
Between our near-IR observations and the original HI detections, an ACS campaign
obtained F606 and F814 images of F575-3 by Cannon \etal (2018).  Using a standard
TRGB distance analysis, Cannon \etal found a distance of 5.11$\pm$0.2 Mpc, slightly
larger than its CosmicFlows-3 distance.

\begin{figure}[!t]
\centering
\includegraphics[scale=0.80,angle=0]{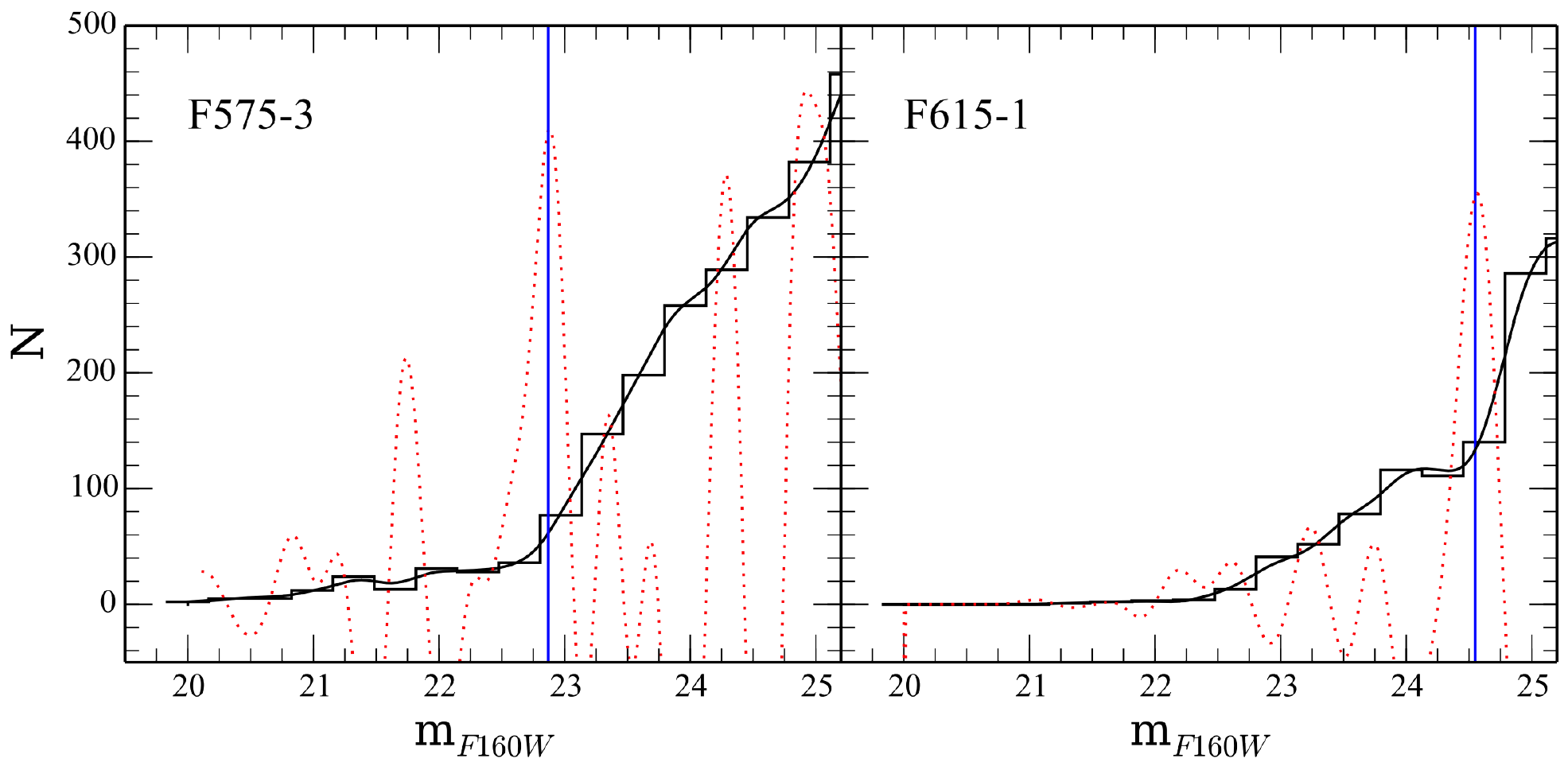}
\caption{\small The luminosity functions for F575-3 and F615-1 above and below the
TRGB (marked as the vertical blue lines).  The normalized histogram ($\sigma$=0.3
mags) is shown as the solid line.  The normalized Sobel filter response is shown as
the red dotted line.  The $m_{TRGB}$ values are 22.87 and 24.55 respectfully with
similar edge errors of $\pm$0.08.
}
\label{sobel}
\end{figure}

While TRGB determination is more difficult in the near-IR bandpasses, we can attempt
a confirmation of F575-3 distance using the F110W-F160W CMD.  From a Sobel filter
test to the RGB (see Figure \ref{sobel}), we find a fit of $m_{TRGB}=22.87\pm0.05$.
A similar analysis for F615-1 finds $m_{TRGB}=24.55\pm0.08$.  Unlike the F814W
filter, the F160W absolute magnitude of the TRGB ($M_{TRGB}$) varies with the mean
color of the RGB (reflecting the mean metallicity of the underlying stellar
population, Freedman \etal 2020).  An empirical correlation was found by Dalcanton
\etal (2012) and updated by Durbin \etal (2020), but we note that this correlation is
about 0.08 mags brighter than expectations from theoretical models and globular
cluster observations.  Using RGB colors (see \S5) of 0.70 for F575-3 and 0.81 for
F615-1, this results in F160W $M_{TRGB}$ values of $-$5.32 and $-$5.58 respectfully
with uncertainties on the mean TRGB color of 0.01 (which translates into an error of
0.02 in the $M_{TRGB}$ calibration).  The resulting TRGB distances are then 4.4 Mpc
for F575-3 and 10.6 Mpc for F615-1.

The F575-3 distance is slightly less than the F814W value from Cannon \etal (2018)
and slightly more than the expected distance from CosmicFlows-3.  Given the more
accurate nature of the F814W TRGB, we adopt the Cannon \etal distance of 5.1 Mpc.
The F615-1 distance is smaller than the expected CosmicFlows-3 distance of 12.1, but
with no other information to draw upon, we adopt a distance of 10.6 Mpc.

\section{RGB Population}

For near-IR CMD's in the literature, the RGB is the most prominent feature and
typically contains over 80\% of the stars brighter than $M_V < -4$.  A distinct upper
edge of the RGB in luminosity is defined by the tip-of-the-RGB (TRGB) effect, where
helium ignition occurs at a fixed temperature and, therefore, constant luminosity
(Sweigart \& Gross 1978).  There are well-known variations in the shape and
brightness of the RGB due to age and metallicity, but most RGB's are dominated by an
old ($\tau > 5$ Gyrs), metal-poor ($[Fe/H] < -$0.5) population.  For many
star-forming dwarfs, there will be contaminant populations of rHeB and EAGB (early
AGB) stars that have similar luminosities and colors as RGB stars (see \S6).
However, the shape and position of the RGB is primarily due to the old stars with
similar metallicities (i.e. during an era of slow enrichment), and the number of EAGB
stars in the RGB region is very small compared to the RGB's themselves.

\begin{figure}
\centering
\includegraphics[scale=0.90,angle=0]{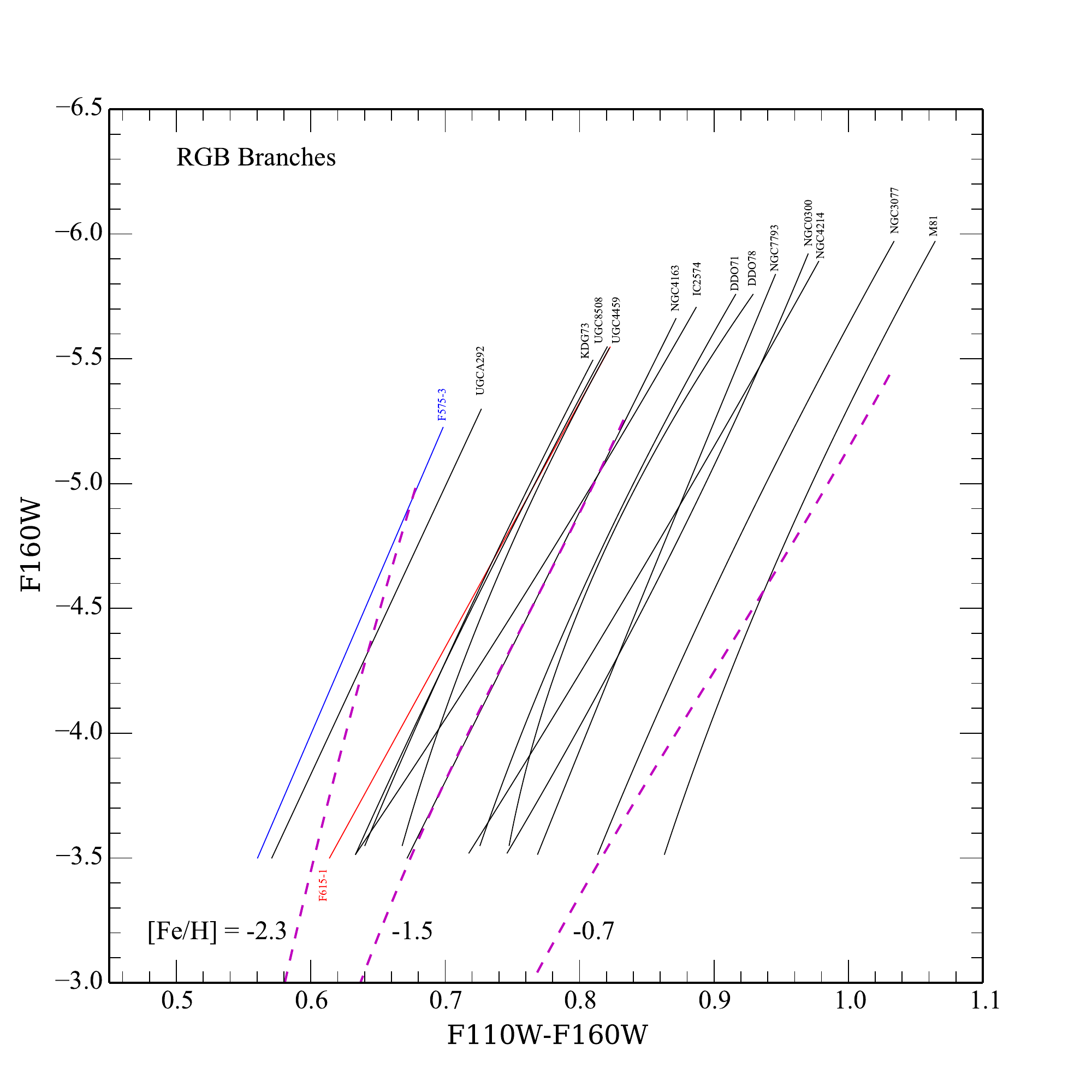}
\caption{\small A comparison of RGB ridgelines from various near-IR CMD's in the
literature.  F575-3 is marked in blue, F615-1 is marked in red.  Three 12 Gyr tracks
for [Fe/H] values of $-$2.3, $-$1.5 and $-$0.7 are also shown.  While the RGB slopes
do not seem to flatten as quickly as the model tracks with metallicity, the general
trend of higher mean metallicity with mean RGB color is clear.  F615-1 displays the
RGB of an intermediate metallicity dwarf ([Fe/H] = $-$1.7); however, F575-3 has an
extremely blue RGB with a mean metallicity comparable to MW globular clusters.
}
\label{rgb}
\end{figure}

The RGB ridgeline tracks for F575-3 and F615-1 (and selected dwarfs from the
literature for comparison) are shown in Figure \ref{rgb}.  While the expectations
from stellar models is that the near-IR RGB is close to a linear vertical feature,
there is some evidence of curvature, particularly for the more massive dwarfs.  For
our analysis, we have taken the shape of the RGB from a ridgeline analysis of the
F110W-F160W CMD between the TRGB and a base limit of $M_{F160W} =-$3.5.  A ridgeline is
preferred for galaxies with strong recent SF as they have a significant rHeB branch
that can dominate the blue side of the RGB.  Selecting the ridge on the red side of
the rHeB selects only those older stars before helium core burning.   

The top of the RGB is defined by the magnitude of the TRGB, which is taken from the
empirical relationship between the color of the top of the RGB and $M_{TRGB}$ (Durbin
\etal 2020).  As can been seen in Figure \ref{rgb}, there is a slight tendency for
more curvature for redder RGB's.  This is easily explained if there is a range of
ages or metallicity to the underlying stellar population, which is expected from the
complex SFH's suggested by optical CMD's (Weisz \etal 2014).  

The centerline of the RGB is a good measure of the global [Fe/H] of a galaxy and a
majority of the high luminosity stars are RGB stars, making the determination of the
centerline computationally easy.  While younger mean age can move the position
slightly blueward, the dominant parameter that determines the centerline of the RGB
is metallicity (parametrized by [Fe/H]).  For example, the difference between a 5 and
12 Gyrs RGB is 0.04 in F110W-F160W, whereas the difference on 0.5 dex in [Fe/H] is 0.16.
We note that this [Fe/H] value will likely under-estimate the
the metallicity of the youngest stars as the
RGB is typically dominated by old stars.  This has been demonstrated by Durbin \etal
(2020) who found the color of the RGB at the TRGB is well correlated with galaxy mass
and, therefore, mean metallicity.  In fact, the red edge of the RGB will basically
define the highest metallicity of the old stars in a galaxy.  

For comparison, three isochrone tracks are shown in Figure \ref{rgb} for a 12 Gyrs
RGB of the indicated $[Fe/H]$ values (Pietrinferni \etal 2006).  As noted by
Dalcanton \etal (2012), the TRGB is significantly higher than predicted by the
isochrones.  The proposed explanation is a mismatch between model colors and HST
filters, but this has not been demonstrated.  The slopes of the RGB's match the
intermediate metallicity RGB isochrones, but are steeper on the red side and too
shallow on the blue side, again, for reasons that are not clear.  

The RGB ridgelines cover a wide range in [Fe/H] from the most metal-rich dwarf
(NGC3077, [Fe/H] $\approx$ $-$0.6) to the smallest dwarf in the ANGST sample, UGCA292
([Fe/H] $\approx$ $-$2.0).  With respect to other near-IR CMD's, F575-3 is on the
extreme blue edge of the samples with a global metallicity that is similar to the
oldest metal-poor MW globulars ([Fe/H] $<$ $-$2.0).  It's TRGB color
(F110W-F160W=0.68) is the bluest of any galaxy with a resolved near-IR CMD.  The blue
edge of the RGB in F575-3 is bluer than any 12 Gyrs isochrone indicating that a
metal-poor and young ($\tau <$ 5 Gyrs) population must also present (see also Cannon
\etal 2018).  

F615-1, on the other hand, is slightly more metal-rich and is grouped with several
other low-mass dwarfs (i.e., KDG73, UGC8508 and UGC4459).  F615-1 sits on the low
side of these low-mass dwarfs with an estimated $[Fe/H]$ value between $-$2
and $-$1.5 (a formal fit produces a value of $-$1.7).  Its RGB is unusual in that it is twice
the width of other dwarf galaxies RGB's (see \S8), so while the mean metallicity is
one of the lowest of the low-mass dwarfs, there is considerable dispersion in the
metallicities of underlying stellar population.

\begin{figure}
\centering
\includegraphics[scale=0.90,angle=0]{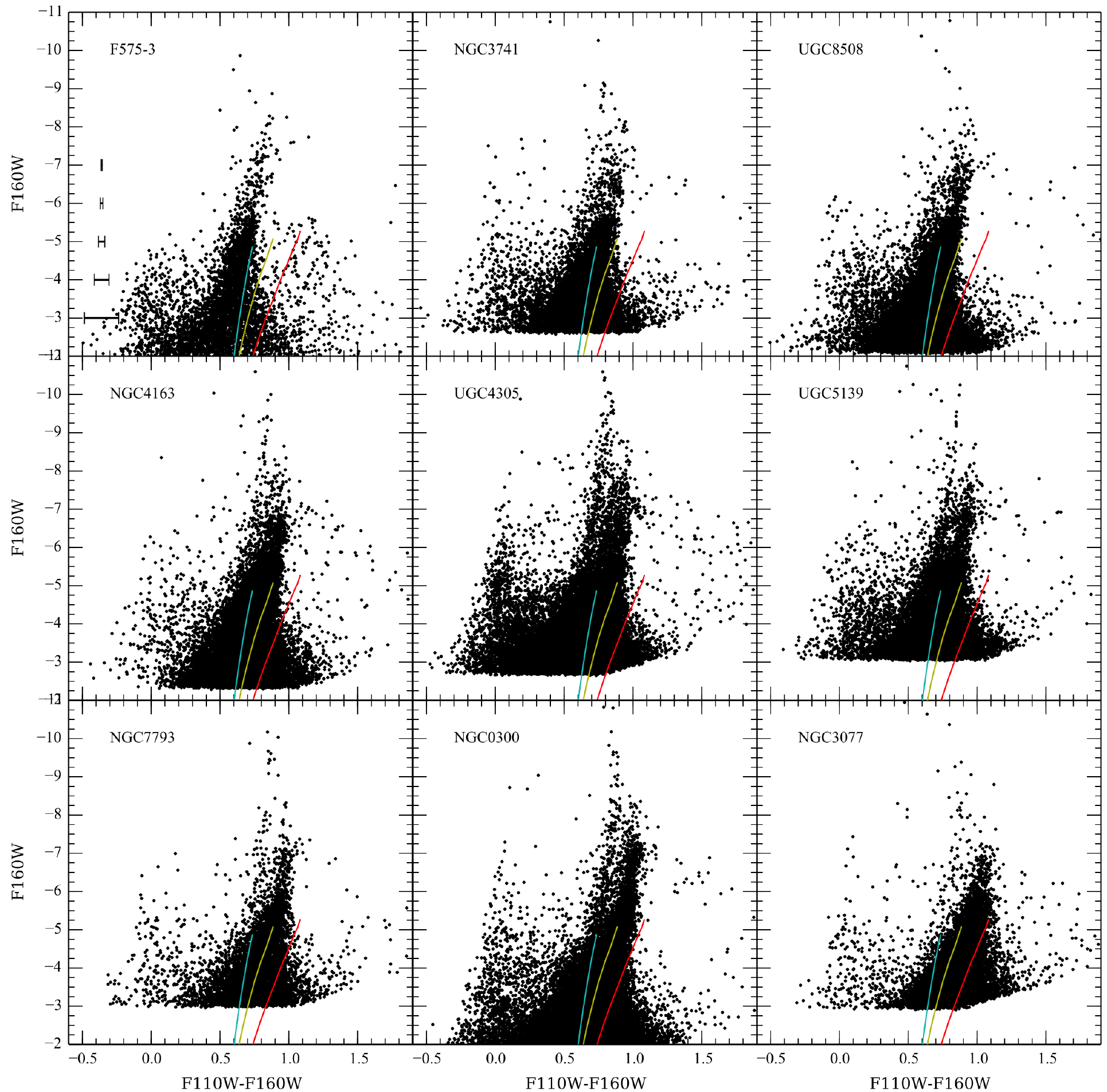}
\caption{\small A comparison of the near-IR CMD's of F575-3 and other young dwarfs in the
literature with similar CMD features (an obvious rHeB, narrow RGB, narrow AGB and a
strong bMS).  Also shown are three 12 Gyr tracks with the metallicities shown in
Figure \ref{rgb} ([Fe/H] = $-$2.3, $-$1.5 and $-$0.7).  Arranged by mean metallicity
from top to bottom, one can see the progression of the red edge of the RGB to higher
metallicities.  The red edge of F575-3 is blueward of even the bluest 12 Gyrs
isochrone indicating a very metal-poor and young RGB population.
}
\label{young_dws}
\end{figure}

\begin{figure}
\centering
\includegraphics[scale=0.90,angle=0]{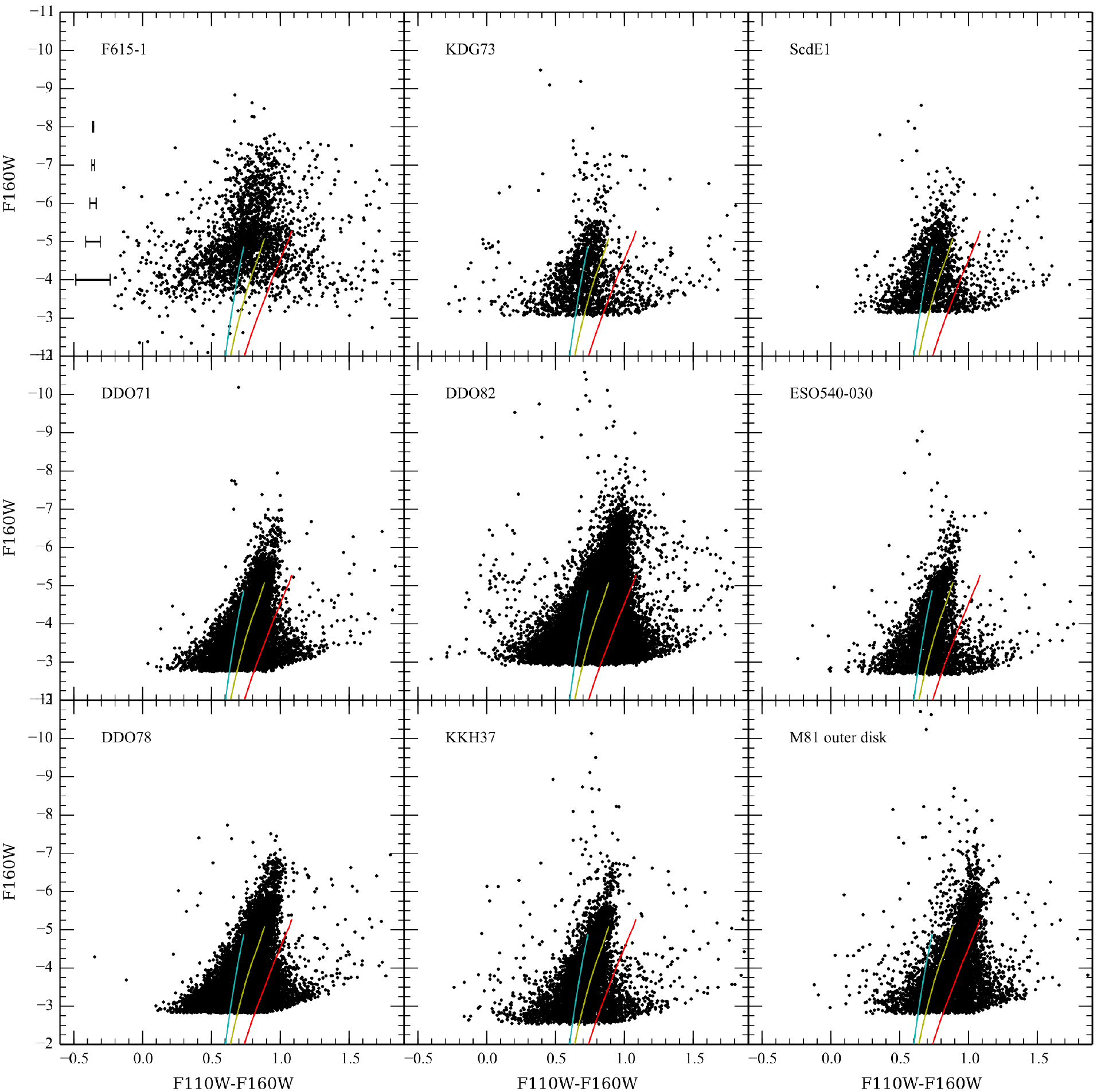}
\caption{\small A comparison of the near-IR CMD's of F615-1 and other old dwarfs in
the literature with similar CMD features (no obvious rHeB feature, few bMS stars and
a weak AGB population).  F615-1 is notable for having an RGB that is twice as wide as
any other old dwarf and a strong AGB population without the luminous AGB's associated
with very recent SF.
}
\label{old_dws}
\end{figure}

A clearer demonstration of the differences in RGB color and shape is found in Figures
\ref{young_dws} and \ref{old_dws}.  Here we have selected a number of near-IR CMD's
from the literature from galaxies of similar stellar masses as F575-3 and F615-1 and
divided the samples into galaxies with prominent RGB's and limited young stars (i.e.,
old) versus those with prominent blue main sequences in the near-IR (i.e., young).
The designation of ``young" versus ``old" by no means minimizes the rich SFH of dwarf
galaxies with episodes of SF that produce a mixture of young and old stars in most
dwarfs.  We simply divide the CMD's by morphology if their features are dominated by
young stellar populations versus older stellar populations.  This is not a statement
of the time of galaxy formation.  Both Figures are sorted by metallicity (from upper
left to lower right) based on the correlation between TRGB color and metallicity from
Dalcanton \etal (2012).

In F575-3, the shape and width of the RGB is similar to other young
dwarf galaxies in the literature.  The narrow RGB in young dwarfs indicates a narrow
range in metallicity and age (i.e., fast chemical enrichment).
The isochrones shown in Figure \ref{young_dws} are all for 12 Gyrs old populations with
[Fe/H] values of $-$2.3, $-$1.5 and $-$0.7.  While most dwarfs have old RGB
populations that match a range in [Fe/H] from $-$1.5 to $-$0.7, F575-3 has the bluest
RGB and only the very lowest metallicity models match the red side of the RGB for a
population older than 5 Gyrs.

Bluer RGB's with higher current metallicities can be matched to the data but only
with younger ages.  For example, at the mean F110W-F160W color of F575-3's RGB, an
age of 1 Gyrs ([Fe/H] = $-$1.5) is required.  This exactly reproduces the scenario
deduced from the optical CMD by Cannon \etal (2018) where over 60\% the stellar mass
of F575-3 is less than 2 Gyrs old.  The surprising caveat is that a population this
young requires the near-IR TRGB to be a full magnitude brighter than observed (in the
near-IR, there is little change in the optically determined F814W TRGB).  In
addition, based on the blue edge of the RGB, the chemical enrichment of the older
population does not appear to have achieved [Fe/H] values above MW globular cluster
metallicities, or the younger population must have formed from newly accreted
metal-poor gas.

The RGB for F615-1 also has some distinct differences when compared to the other
dwarfs with primarily old stellar populations.  The mean metallicity of F615-1's RGB
is similar to other low-mass dwarfs, such as KDG73 and ScdE1.  However, the width of
the RGB is much broader, approximately twice the width at the TRGB than other old
dwarfs in the literature.  While the RGB width in near-IR is primarily a metallicity
effect (i.e., the rapid chemical enrichment from 12 to 5 Gyrs), the broader width in
F615-1 argues for a combination of varying metallicity and varying age in the RGB
population.  For example, the mean metallicity of the F615-1's RGB is approximately
$-$1.7.  This leaves very little room for metallicity variation from initial globular
cluster values of $-$2 in order to widen the RGB.  In fact, the blue side of the RGB
suggests an inverse chemical enrichment scheme.  This region can only be populated
with a young ($\tau <$ 1 Gyrs) population of very low metallicity ([Fe/H $<$ $-$2).
Given that the red side implies a population which is at least older than 5 Gyrs with
[Fe/H] $>$ $-$1, then the younger population must have formed from low metallicity
gas, probably an infall event from extensive reservoir of metal-poor gas as suspected
for F575-3 (see \S8).

\section{Above the RGB}

Stars above the TRGB are comprised from a mixture of origins that, through very
different paths of stellar evolution, end up with high luminosities and RGB colors.
For example, old stars of low metallicity produce a population of thermally pulsing AGB
stars (TP-AGB) near the $M_{TRGB}$ luminosity.  At optical wavelengths, the TP-AGB
stars span a wide range of colors due to a combination of dust in their circumstellar
envelopes and large variations in molecular line spectra with photospheric
temperature (Frogel \etal 1990).  At near-IR wavelengths, AGB stars form into two
distinct groups, those just above the TRGB (having the appearance of an extended RGB,
called M-type AGBs) and those to the red of the RGB (due to the dredge-up of carbon,
referred to as C-type AGBs).   The RGB region below the TRGB has a number of AGB
stars as well (EAGBs); however, their numbers are typically quite small compared to
the number of hydrogen shell burning RGBs in the same region of the CMD (Rosenfield
\etal 2014).  As younger AGB stars are more massive and, therefore, brighter, they
dominate the area above the TRGB.  

Other stars above the TRGB include very young stars (during their helium burning
phase) which form the red helium-burning branch (rHeB) with similar colors to the RGB
but usually higher IR luminosities and typically form a distinct branch on the blue
side of the RGB and AGB sequences.  The TP-AGB population has colors that are
slightly redder than rHeB stars and, except for a small number of extremely red
AGB's, they are indistinguishable in color from RGB stars.  Unless the two sequences
are distinct, the age and metallicity of stars above the RGB can be ambiguous.

A singular fact, from comparison to stellar isochrones, is that a majority the stars
above the TRGB are less than 5 to 6 Gyrs in age and a majority of the stars two
magnitudes above the TRGB (depending on the SFH of the galaxy) are lowest in age of
the AGB population.  Thus, the extent of the AGBs upward in luminosity is a crude
measure of the age of the AGB population as a whole.  The position of the stars above
the AGB, with respect to the RGB ridgeline, is also very sensitive to metallicity.
Given their young age, the metallicity of AGB stars (as given by their colors) are,
effectively, the current [Fe/H] of the galaxy unless there is strong evidence of
significant star formation in the last few Gyrs (i.e., involving more than 50\% the
total stellar mass of the galaxy).

For many of the dwarfs in Figure \ref{young_dws}, the extended RGB and a distinct
rHeB is obvious (e.g., NGC4163 and NGC0300).  With significant recent SF, the rHeB
is populated at higher metallicities than the RGB stars.  Typically, the rHeB
displays a 0.5 to 0.75 increase in [Fe/H] from the ridgeline of the RGB and forms an
excellent measure of current [Fe/H] (see Schombert \& McGaugh 2015).  We note that
F575-3 has a very weak rHeB despite a strong blue main sequence (typically a
signature of SF bursts on scales of 500 Myrs), too weak to estimate the current
[Fe/H] value from the youngest stars. 

To quantify the difference between F575-3 and F615-1 and other dwarfs, we place a box
along the RGB ridgeline ($\pm$ 0.1 in color) from $M_{TRGB}$ to $M_{TRGB}+2$ (as
guided by CMD simulations, this excludes the lower portion of the rHeB) and count the
stars in this box (the RGB fraction).  We then compare that count with a box above
$M_{TRGB}$ but redder than the upper portion of the rHeB (the AGB fraction).  

For the younger dwarfs, the typical RGB/AGB ratio ranges from 8 to 12:1 reflecting
more recent SF (as also indicated by the increase in bMS stars, see \S7).  The ratio
for F575-3 is 9:1, consistent with the other dwarfs that display recent and current
SF.  Though F575-3 displays no current SF, it has the very blue colors of a
star-forming dwarf and an optical CMD that indicates strong SF over the last Gyr.

In general, F615-1's CMD has similar characteristics to other older dwarf CMD's, but
with an apparently stronger AGB population compared to the RGB.  For the older dwarfs
(Figure \ref{old_dws}) the typical ratio for RGB stars to the stars above the TRGB is
between 16 to 20:1.  This reflects the downturn in SF in the last few Gyrs as can be
seen in SFH tracks from Weisz \etal (2014).  There are certainly fewer AGB's with
luminosities brighter than $-$8 in the older dwarfs (DDO82 is the exception with its
relatively strong bMS population).  However, F615-1 has a factor of four more stars
above the TRGB (a RGB/AGB ratio of 4:1) than other old dwarfs.  As this is a
signature of SF in the last few Gyrs, the interpretation is an epoch of higher SF in
the recent past.

A perplexing caveat to these interpretations of the near-IR CMD is that there is no
current SF in F615-1 (no H$\alpha$) and its global colors are similar to other low
SFR systems such as early-type spirals.  There is a mild enhancement of bMS stars
compared to other old dwarfs (see \S7), but none of the bright, ionizing OB complexes
associated with strong SF (see Figure 8 of Schombert \& McGaugh 2015).  In addition,
while there is an overabundance of AGB stars above the TRGB, they are primarily low
in luminosity with very few stars brighter than $-$8.  F615-1 is closest in near-IR
CMD morphology with DDO82 (middle panel of Figure \ref{old_dws}), which both have
small bMS populations.  Often with a young bMS population a number of very bright AGB
stars are found and possibly some rHeB stars (signaling an epoch of star formation in
the last 300 Myrs).  F615-1 is missing this brighter AGB population and any rHeB
population.

The AGB population above TRGB in F615-1 is consistent with a range of ages from 2 to 8
Gyrs, with blue side representing old ([Fe/H] = $-$2) stars and the red side populated
by younger 2 to 5 Gyrs stars and [Fe/H] values reaching to $-$0.5 (see \S6).  We note that
to get the correct ratio of RGB to AGB stars as seen in F615-1's CMD, most of the
stars in the AGB must be between 2 to 6 Gyrs with [Fe/H] values between $-$2 to $-$1.5
suggesting a second epoch of SF using unenriched gas.

We note that our sample does not display an obvious population of carbon-rich AGB
stars in either F575-3 nor F615-1.  In the corresponding $J+H$ ground-based near-IR
filter set, such stars dominate a roughly diagonal sequence that starts near the TRGB
and extends to redder colors and fainter magnitudes (see, for example, the 2MASS
CMD's for LMC in Dalcanton \etal 2012).  There are hints of such C-type AGB sequences
in the CMD's of NGC0300 and UGC4305, but it is not a common feature for all dwarfs.
Part of the motivation for near-IR imaging was the existence of a number of very red
AGB candidates in the F555W-F814W CMD's of LSB galaxies (Schombert \& McGaugh 2014).
However, as can be seen in Figure \ref{be} in \S9, those stars have normal near-IR
RGB colors, i.e., not the colors of carbon-rich AGBs.  

\section{Blue Main Sequence}

The region blueward of F110W-F160W=0.3 is populated by stars representing the most
recent SF ($\tau <$ 100 Myrs), the blue main sequence (bMS).  This region is
dominated by (1) high-mass stars still on the main sequence, (2) stars on the blue
edge of their helium burning phase (bHeB) and (3) stars transitioning between the red
and blue helium burning sequences.  In all of these cases, these are stars with ages
less than 0.5 Gyrs.  The blue stars in this region in the near-IR CMD are also the
same population as is seen in optical CMD's (see Figure \ref{be}).

For the young dwarfs, the typical ratio between RGB stars and the bMS is between 8
and 12:1, although UGC4305 and UGC5139 have ratios near 5:1.  F575-3 has a ratio of
9:1, slightly higher but consistent with other young dwarfs.  For old dwarfs, with
less recent SF, the ratios range from 50 to 200:1.  Surprisingly, F615-1 has a ratio
of 7:1, i.e., an unexpected amount of bMS stars given the red global colors and
no H$\alpha$ emission.  

While the bMS population in both F575-3 and F615-1 is mildly centrally concentrated,
they do not display any star-forming complexes, such as bright OB associations.  This
is in sharp contrast to the bright star-forming complexes seen in our previous HST
imaging of three H$\alpha$ bright LSB galaxies (Schombert \& McGaugh 2015).  
While the lumpy morphology for F575-3 and F615-1 suggested some ancient star groups,
these in fact turned out to be background galaxies.  The underlying stellar
populations in both systems is smooth and devoid of any color coherence reflecting
their star forming past.  This is in agreement with the estimates from the CMD's that
there has been very little SF in the last 500 Myrs, which is sufficient time to mix
the younger and AGB stars through galaxy rotation and turbulence.

There is little to conclude from the bMS population.  The amount of recent SF based
on the bMS is consistent with their stellar mass, as F615-1 has a slightly stronger
bMS population (compared to its RGB) than F575-3.  However, we must reconcile the
lack of current SF, and plentiful gas supply, with the obvious bMS population and
conclude that the recent SFH must have been extremely erratic (i.e., bursts of SF on
timescales of 100 to 500 Myrs) to avoid prominent H$\alpha$ emission but allow
significant numbers of the observed A and B stars.  A skewed IMF would could also be
entertained, but that type of SF is not seen in other LSB galaxies (Schombert \&
McGaugh 2015).

\begin{figure}
\centering
\includegraphics[scale=0.90,angle=0]{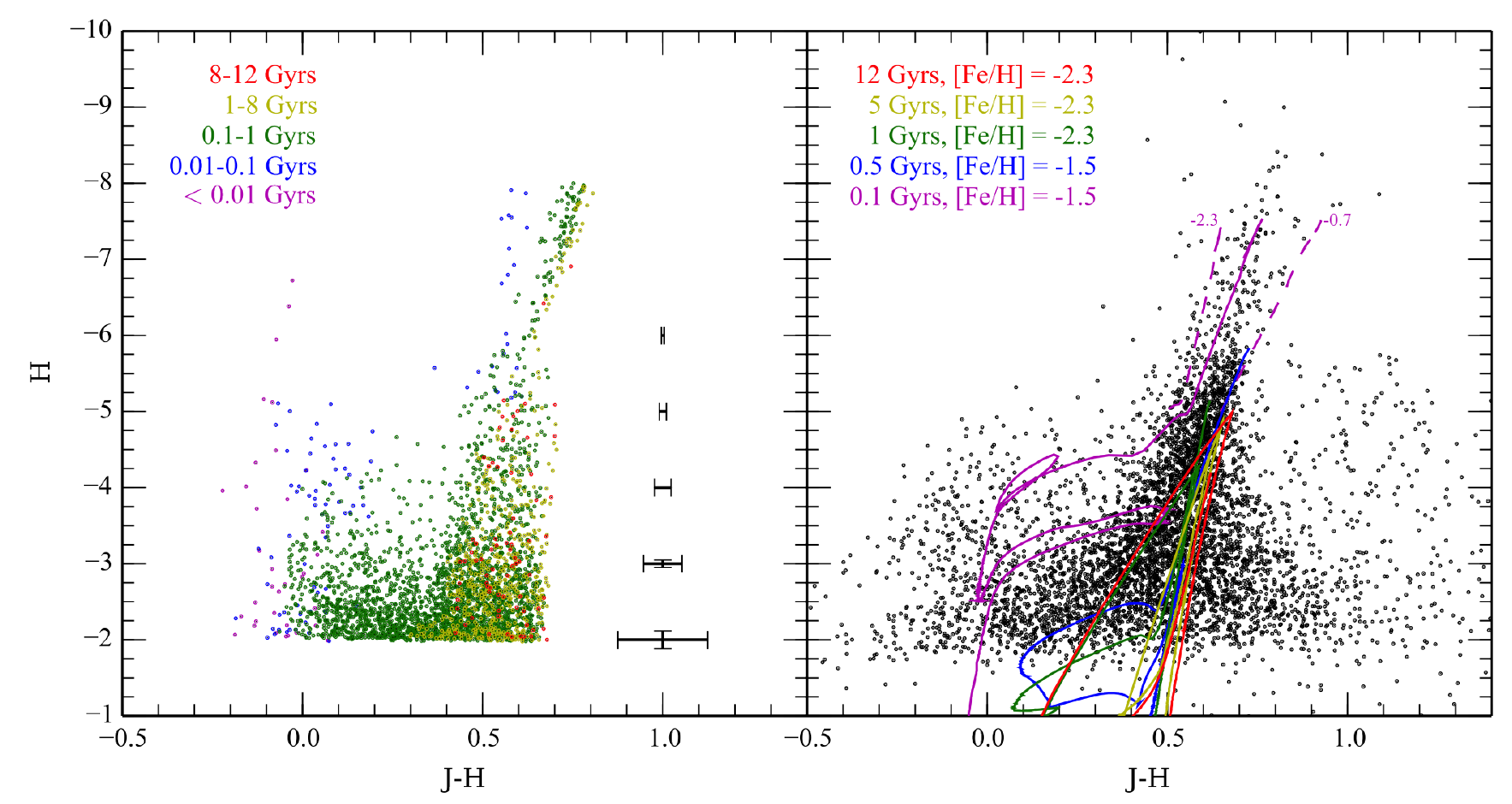}
\caption{\small Isochrone tracks and an IAC-STAR CMD simulation for F575-3.  The
right panel displays the near-IR CMD of F575-3 with a series of simple isochrone
tracks for very low metallicity populations.  The red edge of the RGB constrains the
upper limit on metallicity for populations less than 1 to 2 Gyrs in age.  The mean
color of the AGB region also places tight constraints on the current metallicity of
the galaxy.  The IAC-STAR simulation in the left panel is the near-IR populations for
the SFH proposed from optical CMD's by Cannon \etal (2018), adjusted to the
photometric errors in the data for better comparison with observations.  The IAC-STAR
simulation is an excellent match to the near-IR data (even though deduced from
optical colors).  The only region missing from the data are the massive OB stars that
would produce H$\alpha$ emission, implying a sharp cutoff in SF in the last 100 Myrs.
Missing from the simulation are the stars redward of the RGB, but below the TRGB (see
\S9).
}
\label{tracks1}
\end{figure}

\section{Chemical History}

With a narrow RGB and a well-defined near-IR AGB sequence, it is possible to
reconstruct the chemical enrichment history of F575-3 since its star formation
history (SFH) is well mapped from its optical CMD (Cannon \etal 2018).  In the right
panel of Figure \ref{tracks1} a series of low metallicity ([Fe/H] = $-$2.3)
isochrones are displayed against the RGB data of F575-3 (HST colors have been
converted to $J-H$ for chemical model comparison).  The red edge is a firm limit and
there are no physically real isochrone tracks redward of 0.7 with ages greater than 5
Gyrs and metallicities greater than $-$1.0.  Ages greater than 2 Gyrs only make up
40\% of the underlying stellar population (as deduced from the optical CMD by Cannon
\etal) and the sharp red edge implies very little chemical enrichment up to that age.
Tracks with MW globular metallicities are shown up to a population that is 1 Gyrs old
and are consistent with the width of the RGB.  Thus, we interpret the RGB colors to
imply that no chemical enrichment is required to explain the width of the RGB, a
range of ages outlined by the optical CMD is sufficient.

In addition, the centroid of the AGB region is also extremely sensitivity to
metallicity.  Tracks for a 0.1 and 0.5 Gyrs population with a [Fe/H] value of $-$1.5
are shown (a modest chemical enrichment from MW globular metallicities of $-$2.3)
alongside tracks of $-$2.3 and $-$0.7 for the youngest stars.  The 0.5 Gyrs
population traces the bottom of the AGB region displaying the completing effects of
decreasing age and increasing metallicity.  The metallicity distribution of AGB stars
in F575-3 is consistent with this range of [Fe/H] values for the youngest population.
A 0.1 Gyrs population with a metallicity greater than $-$0.7 would have an RGB color
of 0.8, which is outside the observed RGB boundary.  Thus, we are forced to conclude
that the old stellar population (comprising 40\% of the stellar mass from Cannon
\etal 2018) is composed of metal-poor, MW globular type population of stars.  The
other half of the stellar mass in F575-3 is composed of fairly young ($\tau <$ 2
Gyrs) stars with modest metallicities ($-$1.5 $<$ [Fe/H] $<$ $-$0.5).  Chemical
evolution in F575-3 has been extremely slow, probably at globular cluster levels for
over 10 Gyrs then with a sharp increase to $-$1.5 in only the last Gyr, this late
burst also producing a majority (60\%) of the stars (Cannon \etal 2018).

The left panel of Figure \ref{tracks1} displays an IAC-STAR simulation (Aparicio \&
Gallart 2004) of the SFH from Cannon \etal and the chemical enrichment scenario
deduced above, blurred to the same photometric uncertainty as found in the data.  It
has all of the same features as the data, although F575-3 is missing the stars less than
100 Myrs old (i.e., the bright OB stars what would produce H$\alpha$ emission).
Thus, we conclude some process either 1) halted SF in the last 100 Myrs without
removing a significant fraction of the gas (i.e., not quenching), or 2) the SFH of
F575-3 is one many of microbursts of SF (shorter than 100 Myrs to avoid features in
the near-IR) but sufficient in strength to produce 60\% of the stellar mass over the
last 2 Gyrs.  At the same time, these bursts must only raise the metallicity of the
remaining gas a small amount, a difficult task in a closed box model where even a
small burst results in rapid enrichment.  Otherwise, we conclude later generations
form from a mixture of enriched and non-enriched gas.

\begin{figure}
\centering
\includegraphics[scale=0.90,angle=0]{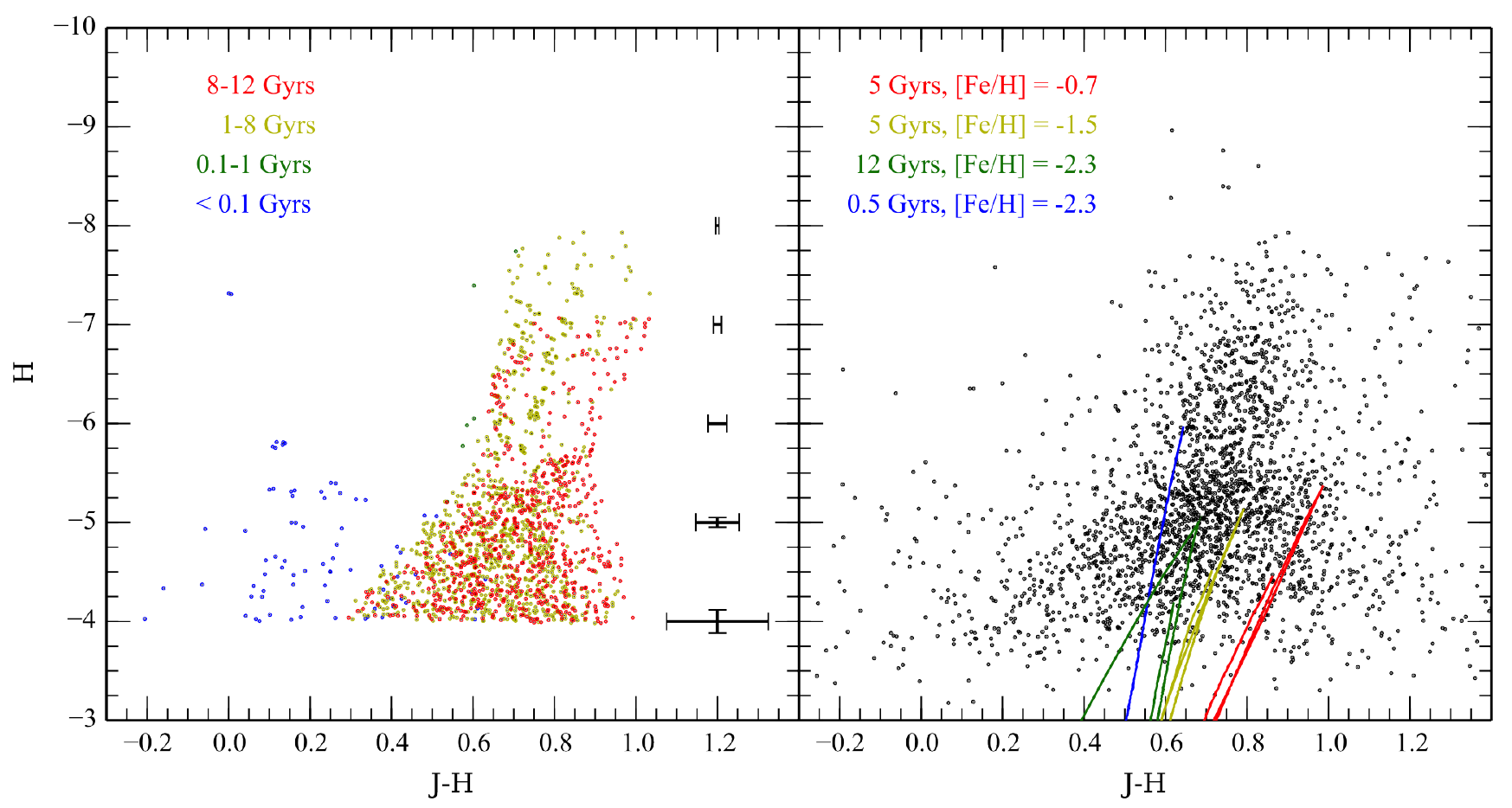}
\caption{\small Isochrone tracks and a CMD simulation for F615-1.  The right panel
displays the near-IR CMD of F615-1 with a series of simple isochrone tracks for very
low metallicity populations.  The red edge of the RGB is limited by a maximal RGB
from a 5 Gyrs population with an intermediate metallicity while the main portion of
the RGB agrees with a 4 to 8 Gyrs population with [Fe/H] values from $-$2.5 to
$-$1.5.  The AGB is also unusually wide, probably due to a range of metallicities.
The IAC-STAR simulation in the left panel is the near-IR populations for a constant
SFR with a small burst at 6 Gyrs to correct for the ratio of RGB to AGB stars (see
text).  The simulation is adjusted to the photometric errors in the data for better
comparison to the observations.  deduced from optical colors).  The only region
missing from the data are the massive OB stars that would produce H$\alpha$ emission,
implying a sharp cutoff in SF in the last 100 Myrs.  Again, missing from the
simulation are the stars redward of the RGB, but below the TRGB (see \S9).
}
\label{tracks2}
\end{figure}

A similar analysis is shown in Figure \ref{tracks2} for F615-1.  Here a series of
younger tracks ($\tau$ = 5 Gyrs) are shown for [Fe/H] values of $-$1.5 and $-$0.7 to
demonstrate the problem of reproducing the red edge of the RGB using low metallicity
values and ages greater than 8 Gyrs.  The blue side is bluer than any low metallicity
([Fe/H] $<$ $-$2) and old ($\tau$ $>$ 8 Gyrs) population can produce.  A young and
metal-poor population (e.g., 0.5 Gyrs and [Fe/H]=$-$2.3 as shown) is required to
reach the blue side of the RGB.

In addition, the AGB region requires a different range of age and metallicities to
produce 1) its red and blue edges plus 2) the lack of very bright AGB's and also 3)
combined with the high proportion of AGB to RGB stars.  A spread of 4 to 8 Gyrs
stars with intermediate metallicities satisfies the above criterion, however, a
constant SF scenario is ruled out for the need of a low metallicity SF burst in the 4
to 6 Gyr age range to increase the AGB to RGB ratio.  As with Figure \ref{tracks1},
an IAC-STAR simulation is found in the left panel of Figure \ref{tracks2}, blurred to
the photometric uncertainties of the data, that include all of the elements discussed
above and a small 500 Myrs population to account for the bMS population.

The uniqueness of the above two population simulations is problematic.  While it is
encouraging that the SFH predicted by the optical CMD in F575-3 is an excellent match
to the near-IR CMD, the lack of current SF drives a microburst interpretation which
is not defined in the simulations.  A similar problem occurs in the simulation for
F615-1 where there is clear evidence for different epochs of star formation with
later SF involving new, metal-poor gas.  The broad width in color of the RGB and AGB
populations argues for a more complicated chemical enrichment history, but the red
and blue edges limit the endpoints to much less than the solar metallicities seen in
high-mass dwarf irregulars.  The mass-metallicity relation can thus be preserved
despite a varying SFH between different types of dwarf galaxies.

\section{Be Stars}

The three major regions of a galaxy's CMD are the RGB, above the RGB (the AGB region)
and the blue main sequence (bMS).  Additional regions, such as the red clump and the
horizontal branch, are only visible in nearby systems.  Each of these three regions
is well explored by standard stellar evolution isochrones and numerous stellar codes
exist to reproduce a CMD from an input star formation history and chemical enrichment
model (see Conroy \& van Dokkum 2016).  Figure \ref{tracks1} and \ref{tracks2} are
our best estimates for the SFH of two dwarf galaxies presented herein based on the
limits outlined in the previous sections.  

However, a fourth region of the CMD is poorly understood or even explored in the CMD
literature.  This is the region below the TRGB and redward of the RGB.  This region
is typically ignored for good reason, as there are no physically meaningful stellar
evolutionary tracks that reach into this region.  As metallicity increases, stellar
isochrones move to the red in near-IR color space.  However, even the tip of an old
12 Gyr population with solar metallicity barely reaches a F110W-F160W color of 1.0.

Inspection of Figures \ref{old_dws} and \ref{young_dws} shows that several dwarfs
have small, but statistically significant, populations of stars with luminosities
below the TRGB, but redder than any possible stellar isochrone (e.g., ScdE1 and
DDO82).  Among the young dwarfs, both UGC4305 and NGC3077 display this behavior.
These stars can not be dismissed as stars with poor photometry as many lie above the
100\% completeness region with errors in color of $\pm$0.05.  Both F575-3 and F615-1
display even higher fractions of stars in this region than other dwarfs in the
literature.  

\begin{figure}
\centering
\includegraphics[scale=0.90,angle=0]{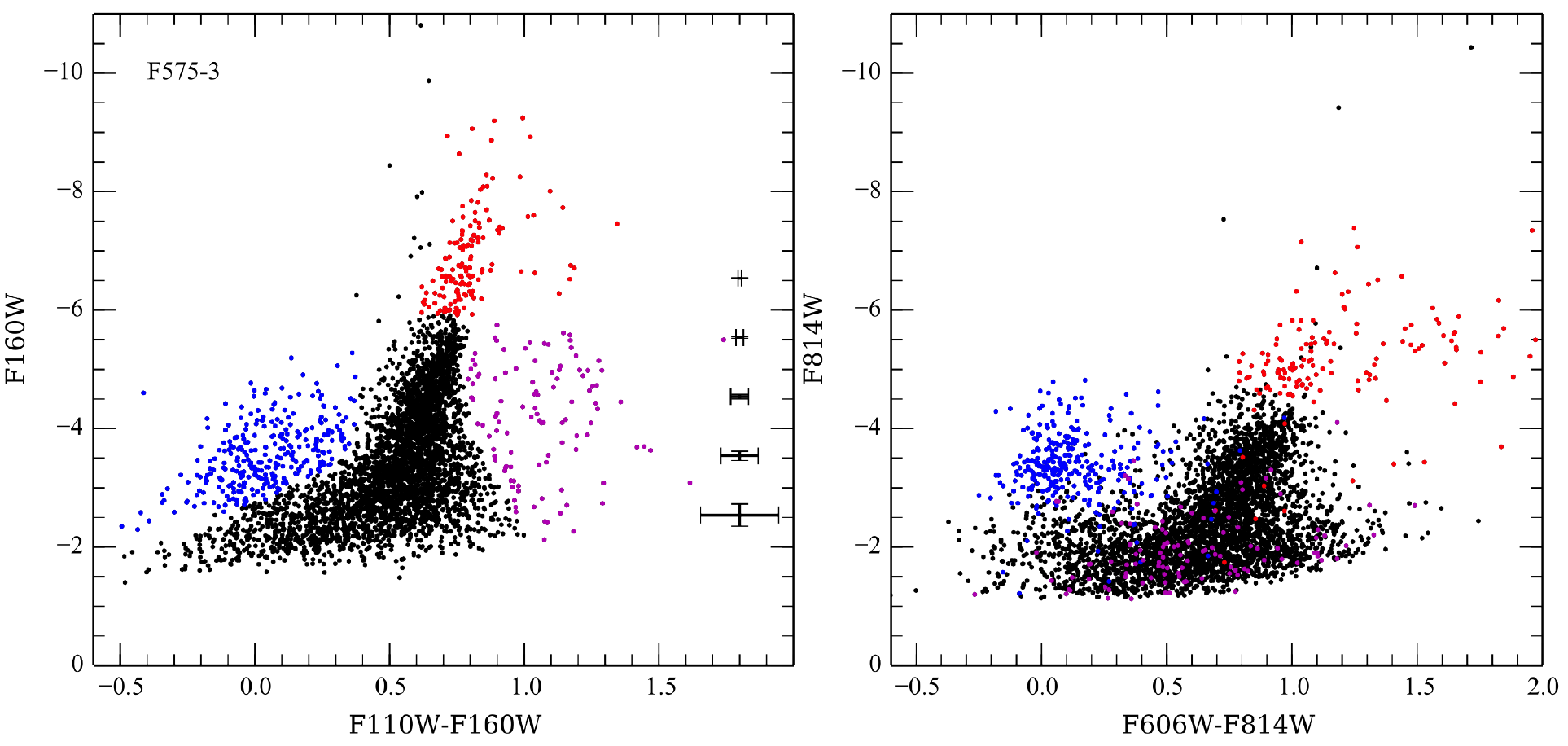}
\caption{\small A one-to-one comparison of stars in the 
the near-IR CMD (F110W-F160W) versus the 
optical CMD (F606W-F814W, Cannon \etal 2018).
Stars in the near-IR portion of the CMD associated with
AGB stars are marked in red, bMS stars in blue and the new Be population marked as
magenta.  The bMS and AGB populations occupy similar regions in the optical and
near-IR, but the Be population are much bluer in the optical than expected from their
near-IR colors.  Stars with an extreme near-IR excess are usually found in systems
with a central post-MS high-mass stars surrounded by a shell of hot gas.  The ionized gas
produces near-IR emission through free-free emission and additional IR luminosity
from scattered light in any associated dust.
}
\label{be}
\end{figure}

An obvious answer to this anomalous population is that they are heavily reddened EAGB
stars.  During the first dredge-up phase it is not uncommon for early AGB stars,
with luminosities below the TRGB, to become shrouded in dust.  Their heavily reddened
colors would occupy this region of the near-IR CMD with F110W-F160W colors between
0.2 and 0.6 mags redder than normal EAGB stars.  This amount of reddening corresponds
to, roughly, between 8 and 15 mags of extinction in $V$ (F555W), so these stars would
be invisible in the optical CMDs.

To check this hypothesis, we can make a direct comparison between this red
population's F110W-F160W colors to their F606W-F814W colors.  For F575-3, F606W/F814W
data were obtained in 2012 HST program GO-12878 and published in Cannon \etal (2018).
For the overlapping field of view, there are 4083 stars in common.  The resulting
F606W-F814W CMD is shown in Figure \ref{be}.  This Figure has been color coded for
its corresponding regions in the near-IR CMD (left panel) such that red symbols are
AGB stars, blue are bMS stars and magenta are stars in this strange region redward of
the RGB.  As can be seen, both the bMS and AGB stars map from optical the near-IR in
similar positions.  In the near-IR, the bMS stars continue to be blue and the AGB
stars are found above the TRGB in the optical plus a few have redder than RGB colors,
probably due to dust shells.  

However, the stars in the red region do not display the characteristics of a heavily
reddened population of EAGB stars (i.e., they should not be detected in the
optical).  Instead, these stars are somewhat fainter with optical colors that place
them between the red side of the bMS and the blue side of the RGB.  In other words,
they have relatively normal optical colors but a sharp, unexpected near-IR excess.

The expectation for foreground MW stars between 21 and 26 $H$ mag is approximately 40
stars in the WFC3 field, most of these being M dwarfs (Robin \etal 2003).
However, the typical M star color is between 0.5 and 0.8 in $J-H$, which is much
bluer than this red population.  In addition, there is no correlation between
Galactic latitude and the occurrence of this red population in other studies.
Based on deep galaxy counts (Cowie \etal 1994), an expected 300 background galaxies
would appear in the field between 21 and 26 $H$ mag.  They would have colors between
0.7 and 1.5 in $J-H$ (based on $B-K$ colors) but 88\% of them would have resolved
diameters greater than 3 WFC3 pixels and be rejected by the photometry algorithms. 
The remaining 40 sources would have red $J-H$ colors, but also red optical colors (their
mean redshifts are around 0.4) and this is not consistent with the optical colors of
the red population (see below and Figure \ref{be}).

\begin{figure}
\centering
\includegraphics[scale=0.90,angle=0]{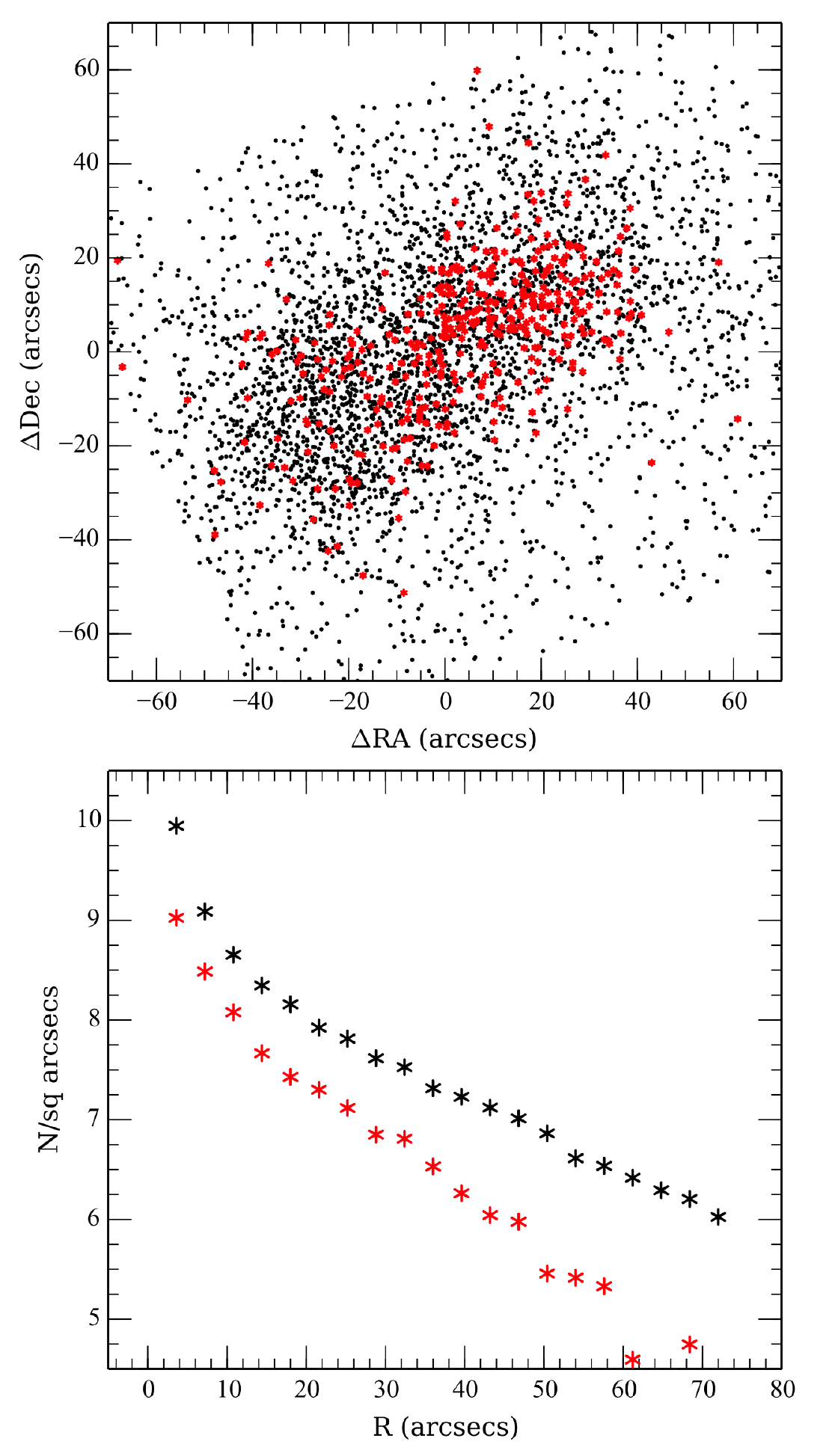}
\caption{\small The distribution of stars brighter than 26 $H$ mags for F575-3
(black) compared to the red Be population (red).  The bottom panel displays the
number of stars per square arcsecs in elliptical annuli from the geometric center.
Note the exponential profile in agreement with the {\it Spitzer} surface brightness
photometry.  The red Be population has a similar shape and slightly lower
scalelength.
}
\label{counts}
\end{figure}

Lastly, if this population was a foreground or background component, then one would
expect it have a spatial distribution of a uniform nature compared to the general
stellar population.  Figure \ref{counts} displays the distribution of star counts for
the total population of F575-3 (black symbols) and the population of stars redder
than $J-H$=0.8.  The red population is clearly as centrally concentrated as the
galaxy stars and does not display a uniform distribution across the field of view as
would be expected for foreground stars or background galaxies.

It turns out that there exists a class of stars with just these very characteristics,
the Ap/Be class of stars (hereafter, just Be stars, Porter \& Rivinius 2003).  Be
stars often display slightly reddened optical colors combined with excess near-IR
luminosity, particularly in the $H$ band ($\Delta(J-H) \approx$ 0.5 to 1.0,
Miroshnichenko \etal 1999).  Based on far-IR photometry and kinematics studies, the
most common explanation for the characteristics of Be stars is that they are
intermediate mass, post-MS stars that formed with a surplus of star-forming debris
(gas and dust).  Due to rapid rotation, the surrounding debris forms into a hot
plasma disk that, when heated by the underlying hot star, produces near-IR excess
due to scattering and free-free emission.  At low metallicities, the dust mass is low
and extinction in the optical is minimal resulting in a luminous star with slightly
redder optical colors and sharply redder near-IR colors.  The near-IR excess is
highly variable due to the rapid evolution of the star and the changing geometry of
the underlying gas cloud, thus, the near-IR excess varies in a manner uncorrelated
with stellar luminosity.  

The properties of Be stars exactly matches the behavior observed for the stellar
population to the red of the RGB.  In addition, as Be stars are a subset of rapidly
evolving high-mass stars, the proportion of stars in the bMS should partially reflect
the number of Be stars on the red side of the RGB.  While it is unclear the duration
of the Be phase, the source of those stars are the ones currently occupying the bMS.
In that regard, there does seem to be a loose connection between the two populations.
For example, NGC3077, NGC3741 and NGC4163 (see Figure \ref{young_dws}) all have
notable Be populations and strong bMS populations.  In the old dwarfs, the few with
recent SF, such as DDO82 and KDG73, also have notable Be populations.  However, there
is no strong correlation between the ratio of the bMS and Be populations nor the
RGB/AGB population.  Our two LSB dwarfs have the strongest Be populations with only
weak bMS populations.  Without a framework or scenario to understand the timescale
for transition from blue star to a peculiar star, the evolution of this population
will await further study.

\section{Conclusions}

Nearly all late type dwarf galaxies reside on the star forming main sequence such
that their current SFR is close to their past average SFR.  This conclusion is based
on the current condition of a galaxy: its total stellar mass and current SFR.  The
path to reach this endpoint can be quite complex, as indicated by the SFH's from
optical CMD's such as Weisz \etal (2014).  In this paper, we investigate the SFH of
two LSB dwarfs whose stellar masses predict detectable current star formation, but
neither have H$\alpha$ emission (Schombert \etal 2013).  HST WFC3 imaging in near-IR
filters produced CMD's with a resolution down to four magnitudes below the TRGB.  We
summarize our observations as the following:

\begin{itemize}

\item{} While both F575-3 and F615-1 are gas-rich, LSB dwarfs, they differ
significantly in their optical and near-IR colors.  Their near-IR CMD's reflect this
same difference in global colors with F575-3 having a very blue RGB and an abundant
bMS population to match its extremely blue optical colors.  F575-3 has similar recent
SF features in its CMD when compared to other star-forming dwarf galaxies in the
literature, yet has no current SF (i.e., H$\alpha$ emission or obvious bright OB
clusters).  F615-1 has a intermediate color RGB and a strong AGB population to match
its early-type optical and near-IR colors.  F615-1 also compares well with dwarfs
that display older population features, but is unusual in having much wider RGB and
AGB sequences and a detectable bMS population.

\item{} The shape and position of RGB in F575-3 indicates that a majority of the old
stars in the RGB are an extremely metal-poor population.  If the oldest stars are
between 8 and 12 Gyrs old (as indicated by the optical CMD in Cannon \etal 2018) then the resulting
[Fe/H] must be at the level of MW globular clusters ([Fe/H] $\approx$ $-$2).  This
would be the most metal-poor stellar population for any galaxy of its total stellar
mass.  The AGB population displays the characteristics of a young (an age between 1
and 2 Gyrs) population with intermediate metallicities ([Fe/H] $\approx$ $-$1.5)
that would make F575-3 a very slow evolving dwarf in terms of chemical enrichment.  

\item{} F615-1 differs from other old dwarfs in having an RGB and AGB region that is
twice as wide as the typical dwarf.  The red side can be reached with a 4 to 6
Gyrs population of intermediate metallicity (an older population with higher
metallicities is inconsistent with the position and numbers of stars in the AGB
region).  The blue side of the RGB requires a population as young, or younger, and
more metal-poor than the population on the red side.  This argues for bursts of star
formation using unprocessed gas to produce the more recent generations of very metal
poor stars.

\item{} The population simulations for F575-3, guided by optical CMD's, are an
excellent match to the observed near-IR CMD and confirm a SFH with a late burst in SF
that produced 60\% of the stellar mass, then halted in the last 100 Myrs (in
agreement with the optical CMD of Cannon \etal 2018).  As the gas supply in F575-3 is high,
quenching is not a solution to the sudden halt in SF.  As the distribution of HI is
fairly uniform (although kinematically disordered), SF appears to be highly erratic,
perhaps induced by recent tidal events (Cannon \etal 2018). 

\item{} The population simulations for F615-1 are less certain as a range of ages,
metallicity and bursts can produce the observed features in the near-IR CMD.  This
lack of uniqueness presents a problem for extracting the exact SFH, however, we can
rule out a simple exponentially declining SF, due to the
mismatch of the AGB to RGB stars.  To reproduce the unusually high AGB to RGB ratios
requires a fairly strong burst approximately 4 to 6 Gyrs ago with low to moderate SF
levels to the present day.

\item{} There is an unusual population of stars in both F575-3 and F615-1 that are
fainter than the TRGB yet impossibly red in the near-IR according to any possible stellar
evolutionary tracks.  These stars have normal optical colors which excludes extinction as
an explanation for their IR excess.  We propose a class of stars analogous to Milky
Way Be stars to explain this population.  As rapidly rotating, optically blue,
post-MS stars, they produce a gas disk that, when heated by the underlying B star,
produces near-IR free-free emission.  This population is clearest in F575-3 thanks to
its very blue RGB, but examination of the CMDs in Figs. 5 and 6 suggests that similar
populations may be president in other dwarfs as well.

\end{itemize}

Despite a slowly evolving SFH for both of these LSB dwarfs, neither match the
definition of ``transitional'' dwarfs as proposed by Dellenbusch \etal (2008).  Both
galaxies are lacking in current SF, but have copious amounts of HI gas and strong
evidence for significant SF in the recent past.  In fact, their positions on the
galaxy main sequence are consistent with other dwarfs of similar stellar mass if one
considers a slightly longer timescale of star formation 
than probed by H$\alpha$ emission.  

With more flexibility to the SFH of a dwarf galaxy, the idea that dwarfs fall into three
basic types becomes more plausible.  These three types, outlined by Weisz \etal
(2015), are 1) early burst, 2) late burst and 3) constant SF.  If SF does not proceed
monotonically, but rather in a series of microbursts, these categories would still
apply under broader SF resolution.  One point to note is that old isochrones ($\tau$
$<$ 8 Gyrs) are nearly identical in the near-IR CMD.  There is no change to our
conclusions if the epoch of initial star formation begins at 12 or 8 Gyrs and very
little difference even at 6 Gyrs.

\section*{Acknowledgements}

This work was based on observations made with the NASA/ESA Hubble Space Telescope,
obtained at the Space Telescope Science Institute, which is operated by the
Association of Universities for Research in Astronomy, Inc., under NASA contract NAS
5-26555.  Support for program GO-15427 was provided by NASA through a grant from the
Space Telescope Science Institute, which is operated by the Association of
Universities for Research in Astronomy, Inc., under NASA contract NAS 5-26555.
Software for this project was developed under NASA's AIRS and ADAP Programs. This
work is based in part on observations made with the Spitzer Space Telescope, which is
operated by the Jet Propulsion Laboratory, California Institute of Technology under a
contract with NASA.  Other aspects of this work were supported in part by NASA ADAP
grant NNX11AF89G and NSF grant AST 0908370. As usual, this research has made use of
the NASA/IPAC Extragalactic Database (NED) which is operated by the Jet Propulsion
Laboratory, California Institute of Technology, under contract with the National
Aeronautics and Space Administration.  One of us [JS] is especially grateful to the
late Karl Rakos, whose conversations of Ap/Be stars in the control room of the KPNO 4m
during long narrow band exposures of distant ellipticals finally reach fruition with
the discovery of these populations in nearby dwarfs.

\end{document}